\documentclass[a4paper,longbibliography,twocolumn,superscriptaddress,pre,showkeys, aps, floatfix,longbibliography ]{revtex4-1}

\usepackage[utf8]{inputenc}
\usepackage[english]{babel}
\usepackage{amssymb}
\usepackage{amsmath}
\usepackage{graphicx}
\usepackage{float}
\usepackage[normalem]{ulem}
\usepackage{color}
\usepackage{multirow}

\usepackage{tikz}

\newcommand{\eqdot}{\,.}
\newcommand{\eqcomma}{\,,}

\usepackage[percent]{overpic}

\begin{document}

\author{Kaj-Kolja Kleineberg}
\email{kkleineberg@ethz.ch}
\affiliation{
Computational Social Science, ETH Zurich, Clausiusstrasse 50, CH-8092 Zurich, Switzerland}
\author{Dirk Helbing}
\affiliation{
Computational Social Science, ETH Zurich, Clausiusstrasse 50, CH-8092 Zurich, Switzerland}

\date{\today}

\newcommand{\piclab}[2]{
\begin{overpic}[width=0.48\linewidth]{#1}
 \put (5,77) {\large \textsf{#2}}
\end{overpic}
}
\newcommand{\piclabb}[2]{
\begin{overpic}[width=0.48\linewidth]{#1}
 \put (5,97) {\large \textsf{#2}}
\end{overpic}
}

\title{Topological enslavement in evolutionary games on correlated multiplex networks}

\begin{abstract}
Governments and enterprises strongly rely on incentives to generate favorable outcomes from social and strategic interactions between individuals. The incentives are usually modeled by payoffs in evolutionary games, such as the prisoner’s dilemma or the harmony game, with imitation dynamics. 
Adjusting the incentives by changing the payoff parameters can favor cooperation, as found in the harmony game, over defection, which prevails in the prisoner's dilemma. Here, we show that this is not always the case if individuals engage in strategic interactions in multiple domains. 
In particular, we investigate evolutionary games on multiplex networks where individuals obtain an aggregate payoff. We explicitly control the strength of degree correlations between nodes in the different layers of the multiplex. 
We find that if the multiplex is composed of many layers and degree correlations are strong, 
the topology of the system enslaves the dynamics and
the final outcome, cooperation or defection, becomes independent of the payoff parameters.
The fate of the system is then determined by the initial conditions.
\end{abstract}


\maketitle

\section{Introduction}

Strategic interactions between individuals are at the root of society.
It is highly beneficial to a population if these interactions lead to large-scale cooperation~\cite{SuperCooperators,Smith:transitions}.
A widely used approach to promote cooperation relies on economic incentives. 
Examples include performance based bonuses in enterprises, reduced costs for the disposal of sorted waste to encourage recycling, or sharing the financial impact of climate change~\cite{Milinski2008}.
Large scale cooperation can emerge through individual
strategic interactions that are commonly described by evolutionary game theory~\cite{nowak06evolutionaryDynamicsBOOK,smith:evo,Axelrod1984}, where incentives are incorporated into payoffs that individuals earn from playing games with their neighbors in a network of contacts~\cite{Nowak1992,Perc2013,Traulsen2006,Rand2014,Pinheiro2012}.
The impact of structured populations in a single domain on the evolution of cooperation is well understood~\cite{Santos2005,GmezGardees2007,PERC20171,metric:selection,collective:navigation}.
However, in reality, interactions take place in different domains, such as business, circles of friends, family etc. 
Such interactions can be captured by multiplex networks, which are systems comprised of several network layers, where the same set of individuals are present~\cite{ginestra:natphys,arenas:multiplex,pre:multiplex:correlations,geometry:multilayer,geo:targeted}. 
The impact of multiplexity on the outcome of evolutionary games is of high importance for the emergence and stability of cooperation in real systems and has recently attracted a lot of attention~\cite{Santos2014,Wang2012,PhysRevE.86.056113,Wang2015,perc:multiplex:public,Wang2012,Wang2013,Jiang2013,Szolnoki2013,PhysRevE.89.052813,jesus:mutliplex_game,perc:neutral,perc:optimal:inter,gamesoc}.
Especially the interplay between the structural organization of the different domains in the multiplex---for example whether there are correlations between the importance of an individual in the business and social domain---and evolutionary game dynamics is still not well understood.

In this paper, we show that 
the interplay between evolutionary dynamics and the structural organization of the multiplex
can have dramatic consequences for the effectiveness of incentive schemes.
In particular, if the degree of nodes (which may abstract their importance) is correlated among different domains, 
which is the case in most real multiplexes~\cite{pre:multiplex:correlations,pre:degree:correlations:2,prl:degree:correlations,scirep:degree:corr,geometry:multilayer,geo:targeted},
the evolutionary dynamics can become enslaved by the topology (\textit{topological enslavement}). 
This means that the hubs may dominate the game dynamics.
In this case the ability of incentives to control the system breaks down completely and the outcome becomes independent of the payoffs.
This phenomenon can be interpreted as an encrusted society which is insensitive to incentives.

\section{Results}

Social and strategic interactions naturally occur in different domains, such as business, friends, family etc. 
Each of the interaction domains is given by a network of contacts. These networks are usually 
heterogeneous---often scale-free---as well as highly clustered~\cite{newman:structure}, i.e. they have a large number of closed triangles~\cite{Newman2010}.
The simultaneous presence of individuals in different domains is naturally represented by a multiplex network~\cite{Boccaletti2014}.
A multiplex is comprised of several network layers, each of which consists of the same set of nodes. 
In real multiplexes, the topologies of the different layers are not independent from each other. 
Specifically, real multiplex networks have been shown to be far from random superpositions of their constituent layer topologies.
Instead, they exhibit a large number of overlapping edges~\cite{Boccaletti2014}, the degrees of nodes between different layers are correlated~\cite{pre:multiplex:correlations,pre:degree:correlations:2,prl:degree:correlations}, and nodes tend to connect to similar nodes in different layers~\cite{geometry:multilayer} (similarity correlations).

Let us first focus on the effect of degree correlations. We construct correlated and uncorrelated multiplexes using Barabasi-Albert (BA) networks in the individual layers. From these networks, we construct multiplexes either by randomly matching individuals between the two layers, in this case there are no degree correlations, or by matching individuals according to their degree rank, which leads to maximal degree correlations. 
We now simulate the evolutionary game dynamics. 
Individuals play games with their contacts in each layer of the multiplex. In each layer, individuals have two strategic choices: they can either cooperate (C) or defect (D). The payoff of each two-player game is then described by the payoff matrix
\begin{equation}
M=
\begin{array}{c|cc}
\quad & \text{C} & \text{D} \\
\hline
\text{C} & 1 & S \\
\text{D} &  T & 0 
\end{array} \eqdot
\label{eqn_payoff_matrix}
\end{equation}
Parameters $T$ and $S$ define different games~\cite{nowak06evolutionaryDynamicsBOOK}. $T<1$ and $S>0$ defines the ``harmony'' game, $T<1$ and $S<0$ corresponds to the ``stag hunt'' game, $T>1$ and $S<0$ yields the ``prisoner's dilemma'', and finally for $T>1$ and $S>0$ we obtain the ``snowdrift'' game.
One round of the game consists of each individual playing one game with each of her neighbors in each layer in the multiplex. 
Furthermore, we consider the evolution of the system to be governed by imitation dynamics~\cite{Szab2007}, reflecting that individuals tend to adopt the strategy of more successful neighbors. After each round of the game (synchronized updates) each node $i$
chooses first one layer, $l$, at random and then---within this layer---one neighbor $j$ at random, and copies her strategy with probability $P_{l,i \leftarrow j}$, specified by the Fermi-Dirac distribution~\cite{Szab2007} (in analogy to maximum entropy considerations in Glauber dynamics)
\begin{equation}
 P_{l,i \leftarrow j} = \frac{1}{1+e^{-(\Pi_{j} - \Pi_{i})/K}} \eqdot
\end{equation}
Herein,
\begin{equation}
 \Pi_{i} = \frac{1}{n_l} \sum_{l=1}^{n_l} \pi_{i,l}
 \label{eqn:Pi}
\end{equation}
measures the aggregated payoff of node $i$ given by the sum of the payoffs $\pi_{i,l}$ of node $i$ in layer $l$ over all layers, which we normalize by the number of layers. Parameter $K$ plays the role of a temperature and quantifies the irrationality of the players. In the Supplementary Materials, we show that different update rules yield qualitatively similar results.
After all nodes updated their strategy simultaneously, we reset all payoffs. In this paper, if not stated otherwise, we always start with $50\%$ cooperators, which are randomly assigned. Later, we will investigate and discuss the impact of different initial fractions of cooperators.
As order parameter, we use the mean final cooperation as 
\begin{equation}
 c = \frac{1}{n_l} \sum_{l=1}^{n_l} c_l \eqcomma
\end{equation}
where $c_l$ denotes the final mean cooperation in layer $l$, and $n_l$ is the number of layers.

\begin{figure}[t]
\includegraphics[width=1\linewidth]{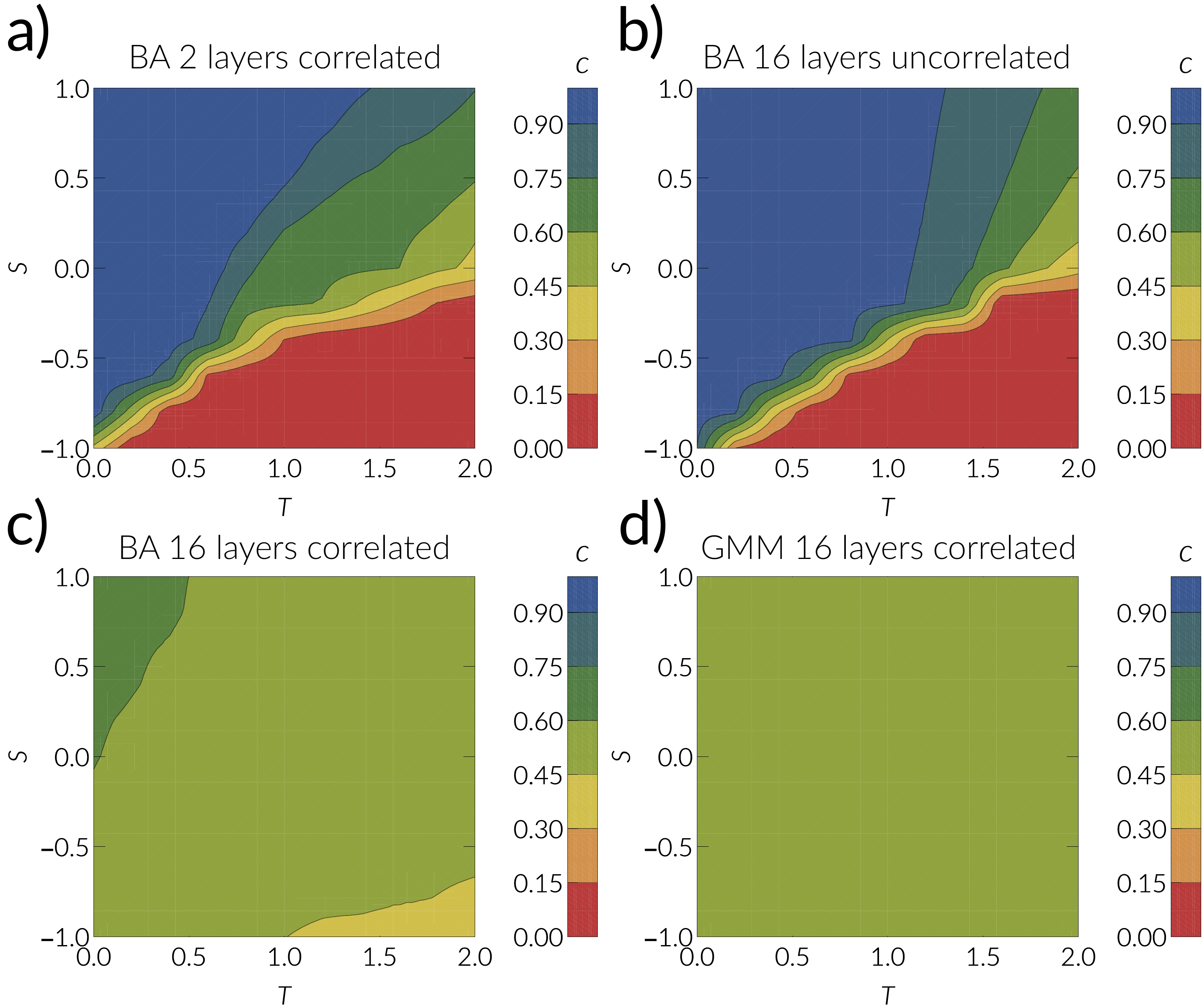}
 \caption{
 Mean final cooperation (color coded) as a function of the game payoff parameters $T$ and $S$.
 Layers have always $N=4000$ nodes. Results are averaged over $400$ realizations.
 \textbf{a)} $2$ layers BA multiplex with degree correlations. The case for $2$ layers without degree correlations in qualitatively similar and shown in Supplementary Figure 3.
 \textbf{b)} $16$ layers BA multiplex without degree correlations. 
 \textbf{c)} $16$ layers BA multiplex with degree correlations.
 \textbf{d)} $16$ layers GMM multiplex with degree correlations ($\nu=1$), no similarity correlations ($g=0$), power-law exponent $\gamma = 2.6$, and mean local clustering coefficient $0.4$. 
 \label{fig:st}}
\end{figure}

The results show that a sufficiently large number of layers and the existence of degree correlations give rise to a particularly interesting phenomenon. If the number of layers is low we obtain a qualitatively similar behavior as compared to the outcome in single networks, see Fig.~\ref{fig:st}a. The results for a single network are shown in the Supplementary Materials. In the absence of degree correlations, increasing the number of layers only leads to mild changes, cf. Fig.~\ref{fig:st}b. However, 
if degree correlations are present and the number of layers is large enough, 
in the whole $T-S$ parameter space considered we now observe a mean final cooperation of $c \approx 0.5$, which is nearly independent of the game payoff parameters $T$ and $S$ (see Fig.~\ref{fig:st}c). 
We now consider more realistic multiplexes, for which we use the geometric multiplex model (GMM) developed in~\cite{geometry:multilayer}. This model has several advantages: we can tune the degree of heterogeneity of the layer topologies (power-law exponent $\gamma$), we can generate networks with realistic mean local clustering coefficients, we can tune the strength of degree correlations between the layers by varying parameter $\nu \in [0,1]$ ($0$ means no degree correlations, and $1$ maximal correlations), and finally we can also tune similarity correlations that have been found to exist in real multiplexes by varying parameter $g \in [0,1]$ (the combination of these correlations controls the amount of overlapping edges, another important property of real multiplexes). 
As shown in Fig.~\ref{fig:st}d, using these more realistic multiplexes yields similar results as discussed before. Interestingly, the behavior does not change significantly as one varies the overlap (or parameter $g$ respectively, see Supplementary Materials).  
These findings suggest that degree correlations are responsible for the observed phenomenon. 

\begin{figure}[t]
 \includegraphics[width=1\linewidth]{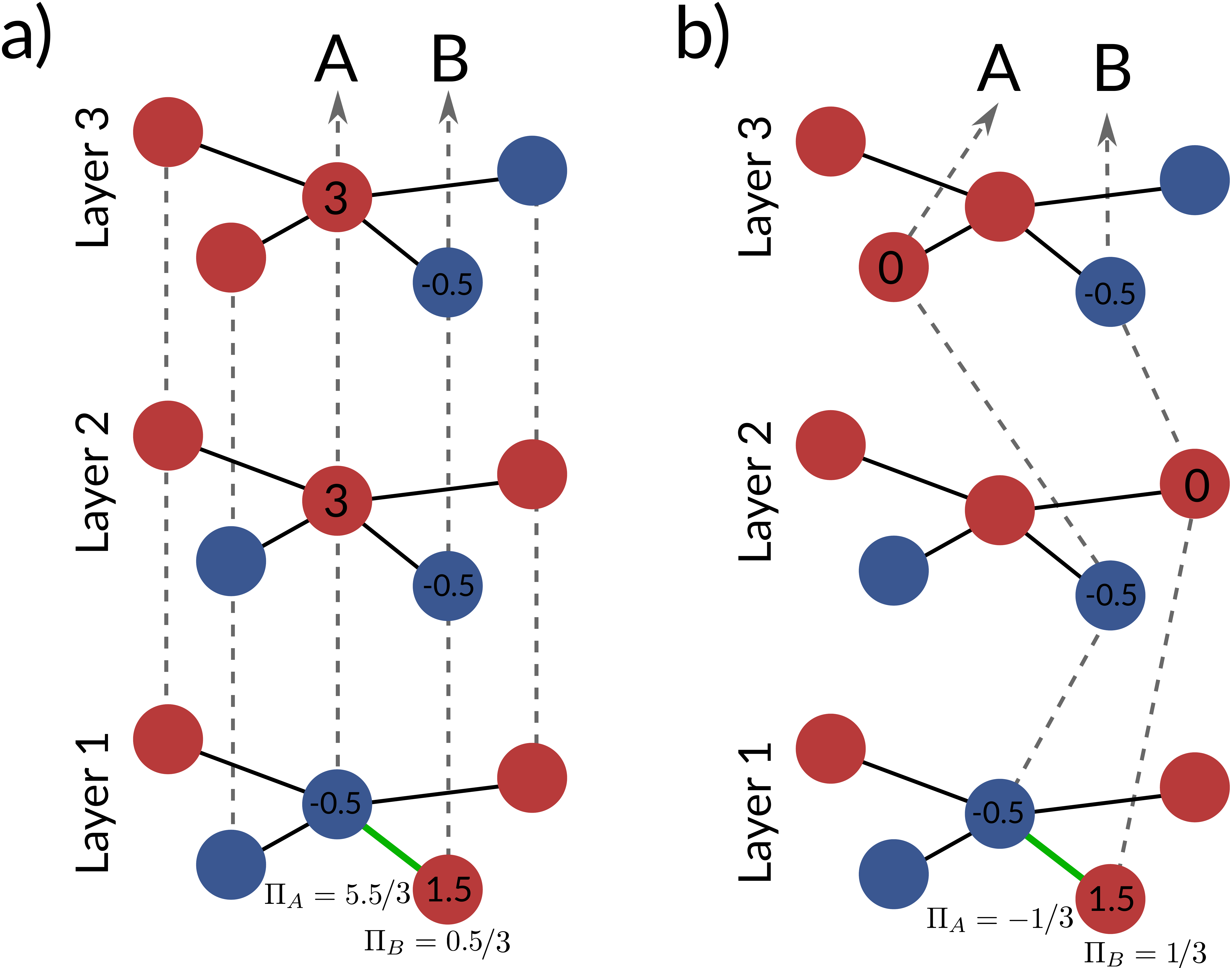}
 \caption{ 
 A simple toy example of a three layer multiplex with one hub and four leave nodes to illustrate topological enslavement. Red denotes defectors and blue cooperators. 
 Payoffs that selected nodes have earned in the prisoner's dilemma ($T=1.5$ and $S=-0.5$) in each layer are displayed on top of the nodes. 
 Dashed lines connect the same node in different layers. 
 \textbf{a)} Strong degree correlations are present, hence node A is the hub in all three layers simultaneously.
 \textbf{b)} No degree correlations. Node A is the hub in layer $1$, but a leave node in layers $2$ and $3$.
 Here we omitt the dashed lines for all nodes except A and B for better readability. 
 \label{fig:sketch}
 }
\end{figure}

The mechanism at play is ``topological enslavement''. Individuals only have knowledge of the aggregated payoff but imitate the strategy of other individuals in a particular interaction domain. This imitation process is blind to where the actual payoff came from, which could have been earned in another layer following a different strategy.
Hubs have the potential to earn higher payoffs because they play more games. Furthermore, due to their high number of links, nodes are more likely to select hubs as imitation candidates. 
If degrees are correlated between the layers, hubs in one layer are also hubs in another layer. 
Let us consider the harmony game with $S=0.5, T=0.5$ as an example. 
Assume that we have a hub that is initiated as cooperator in layer $1$ and defector in layer $2$. Then, in the first round, in layer $1$ the hub earns an expected payoff of $k/2 \cdot 1.5$ (recall that we assign to each node the initial state of cooperate with $50\%$ probability). In layer $2$ it earns $k/2 \cdot 0.5$, which 
yields an aggregated payoff of $k/2$ according to Eq.~\eqref{eqn:Pi}. 
For the sake of simplicity, assume that we have, apart from the hub, in each layer $k$ other nodes that are exclusively connected to the hub. These nodes then earn a payoff in the range $[0.5,1.5]$.  
Now, if $k \gg 1$, the other nodes will always imitate the hub, and not vice versa, and hence in layer $2$ where the hub started as a defector eventually defection will prevail, although this could not happen in the harmony game in isolation. 
As we have seen in this toy example, hubs can accumulate a high aggregated payoff, and if degree correlations are present, the topology dominates the game dynamics. The consequence is that the actual game payoffs $T$ and $S$ become irrelevant~\cite{gru2014}. Instead, the outcome is determined by the initial conditions. 
The behavior is similar in the prisoner's dilemma game.  
In Fig.~\ref{fig:sketch}a we present a toy example to illustrate the mechanism for $T=1.5$ and $S=-0.5$. 
Let us focus on the green link in the figure connecting nodes A and B in layer $1$. In this layer, node A cooperates and earns a payoff of $1 \cdot 1 + 3 \cdot S = -0.5$. Node B defects and earns a payoff of $1 \cdot T = 1.5$. 
However, the aggregate payoffs as defined in Eq.~\ref{eqn:Pi} are $\Pi_A = 5.5/3$ and $\Pi_B=0.5/3$. Hence, even if the defector B earns a higher payoff than A in layer $1$, the aggregate payoff of A is significantly higher and B is likely to imitate the cooperating hub.
This is due to the multiplex topology, in particular the existence of degree correlations (A is a hub in all layers).
In other words, especially in the early rounds of the game, the high payoff a hub earns in layers where it defects makes its strategy worth imitating even in layers where it cooperates. This effect leads to topological enslavement, where the initial state of the hub determines the outcome of the entire layer (we show this with numerical simulations in the following). 
Let us now consider the case without degree correlations as illustrated in Fig.~\ref{fig:sketch}b. 
Node A is still a hub in layer $1$, but a leave node in the two other layers. In the example presented in the figure, the aggregated payoffs of nodes A and B are $\Pi_A = -1/3$ and $\Pi_B=1/3$ respectively. Hence, it is likely that node A, the hub, switches to defection and subsequently defection will spread across the system. In this case, the hub is not able to drive the entire layer to cooperation, and topological enslavement is not observed.

\begin{figure}[t]
  \includegraphics[width=1\linewidth]{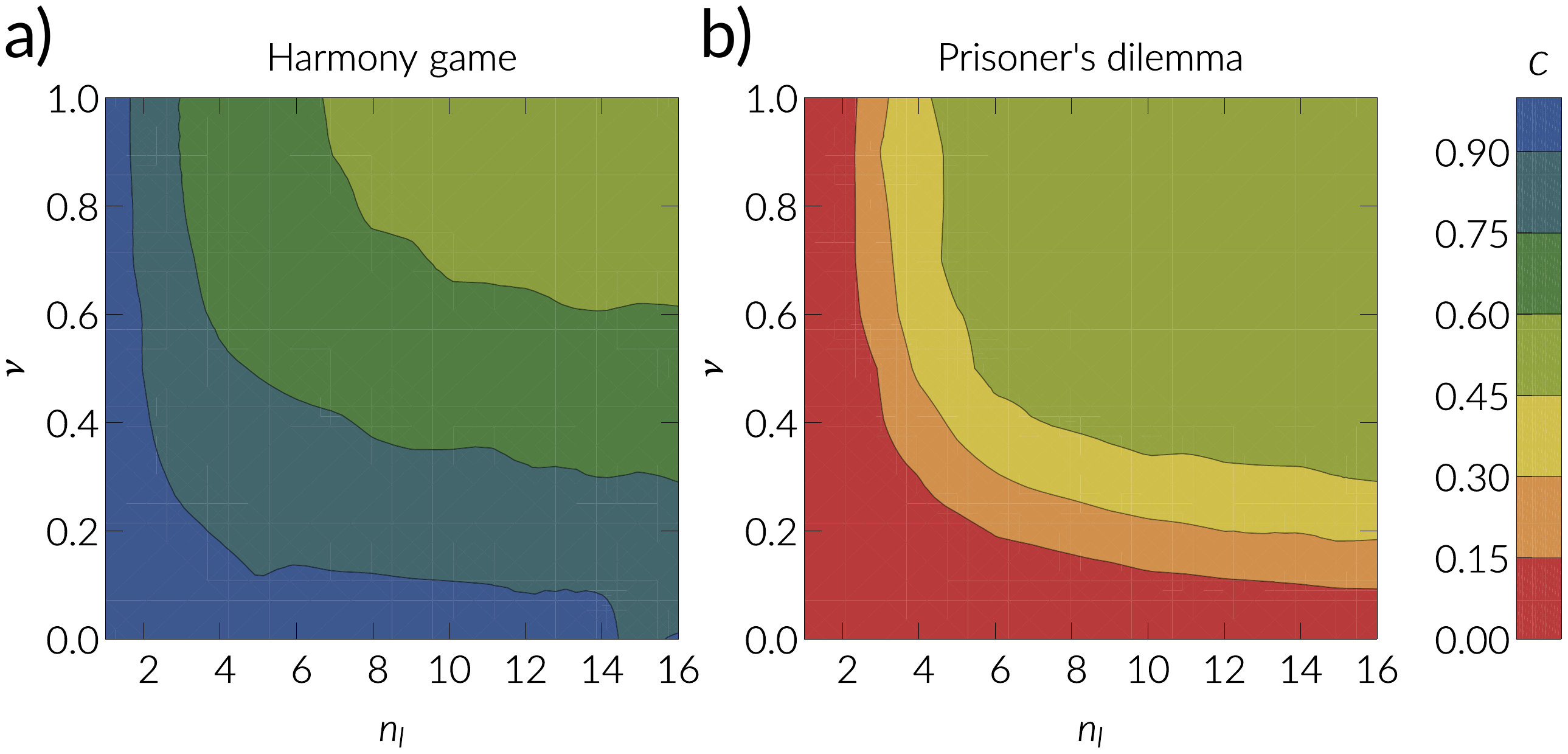}
 \caption{
  Mean final cooperation (color coded) for \textbf{(a)} the harmony ($T=0.5$ and $S=0.5$) game and \textbf{(b)} the prisoner's dilemma ($T=1.5$ and $S=-0.5$)
  as a function of the number of layers $n_l$ and the strength of radial correlations $\nu$.
   Networks are generated with the geometric multiplex model described in the text. Layers have $N=4000$ nodes, power-law exponents $\gamma=2.6$, and mean local clustering coefficient $0.4$. 
   Results are averaged over $400$ realizations.
 \label{fig:NlNu}}
\end{figure}

\begin{figure}[t]
  \includegraphics[width=1\linewidth]{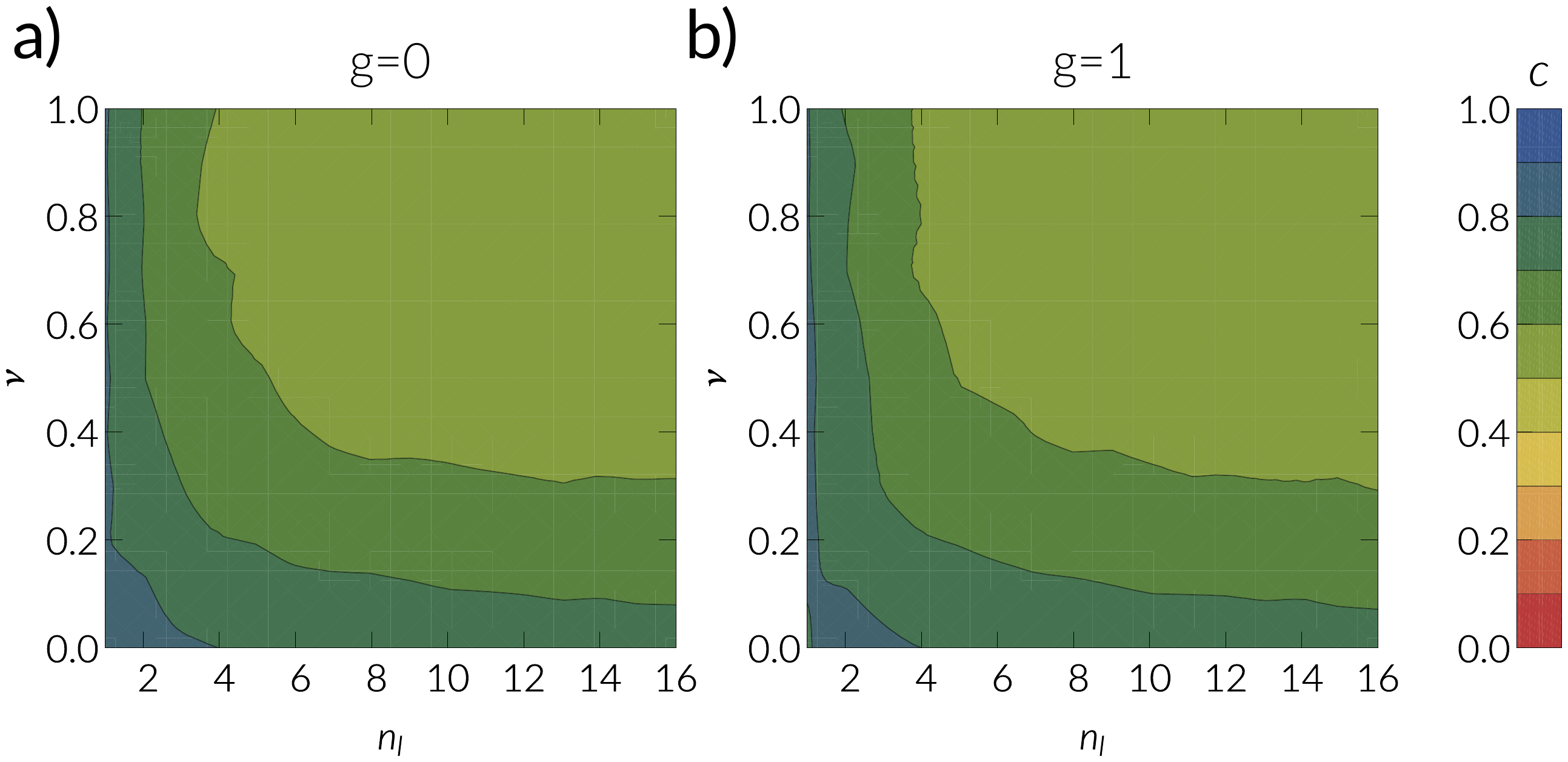}
\caption{Mean final cooperation (color coded) in the public goods game as a function of the number of layers $n_l$ and the strength of degree correlations $\nu$. 
 Networks are generated with the geometric multiplex model. Layers have $N=4000$ nodes, power-law exponents $\gamma=2.5$, mean local clustering coefficient $0.4$.
 The synergy factor is $r=1.6$. Results are averaged over $400$ realizations
 \textbf{a)} No similarity correlations ($g=0$). 
 \textbf{b)} With similarity correlations ($g=1$).
 \label{fig:pgg}}
\end{figure}

Topological enslavement emerges and triggers payoff irrelevance as we increase the number of layers and the strength of degree correlations.
In a single domain, defection is the prevailing strategy in the prisoner's dilemma and cooperation flourishes in the harmony game. This also applies to the topologies considered here, see Supplementary Materials.
In the following we choose these games because they are prototypes of very different situations, where the outcome in a single domain is either cooperation or defection.
However, even in these complementary cases, we find that if the number of layers is large enough and degrees correlations are strong enough,
topological enslavement makes the final outcome of these two games indistinguishable. 
This is shown in Fig.~\ref{fig:NlNu}. If degree correlations are weak (small $\nu$) and/or the number of layers is small (low $n_l$), we recover the known result from a single domain, namely cooperation in the harmony game and defection in the prisoner's dilemma. 
However, if degree correlations are strong enough and the number of layers is large enough, both games show the same outcome: a mean cooperation of $c \approx 0.5$ (see green area in Fig.~\ref{fig:NlNu}).

Finally, topological enslavement is not restricted to pairwise games like the prisoner's dilemma or the harmony game. Let us consider the public goods game as an example, where individuals play in different overlapping groups. A node with degree $k$ participates in $k+1$ groups centered around each of its neighbors and itself~\cite{san08}. Cooperators contribute to the common pool with a total amount that they divide equally among the groups they participate in. 
The total amount in the common pool of each group is multiplied by a factor $r$ and distributed equally among all members of the group. The game is played independently in different layers, and the total payoffs are aggregated for each node (see Supplementary Materials for details).
The results are shown in Fig.~\ref{fig:pgg}. For the given synergy factor, cooperation density is large in the absence of degree correlations, ($>70\%$), but if the number of layers is large and degree correlations are present, cooperation density drops to a value close to $50\%$ (independently of parameter $g$ and hence of the overlap in the multiplex. The behavior is insensitive to a change of parameter $g$, and hence of the overlap in the multiplex, in contrast to recent findings in lattice topologies~\cite{perc:multiplex:public}. In other words, the effect of heterogeneity and degree correlations is the dominant factor).
These findings suggest that topological enslavement can occur in a broad range of games.

\begin{figure}
 \includegraphics[width=1\linewidth]{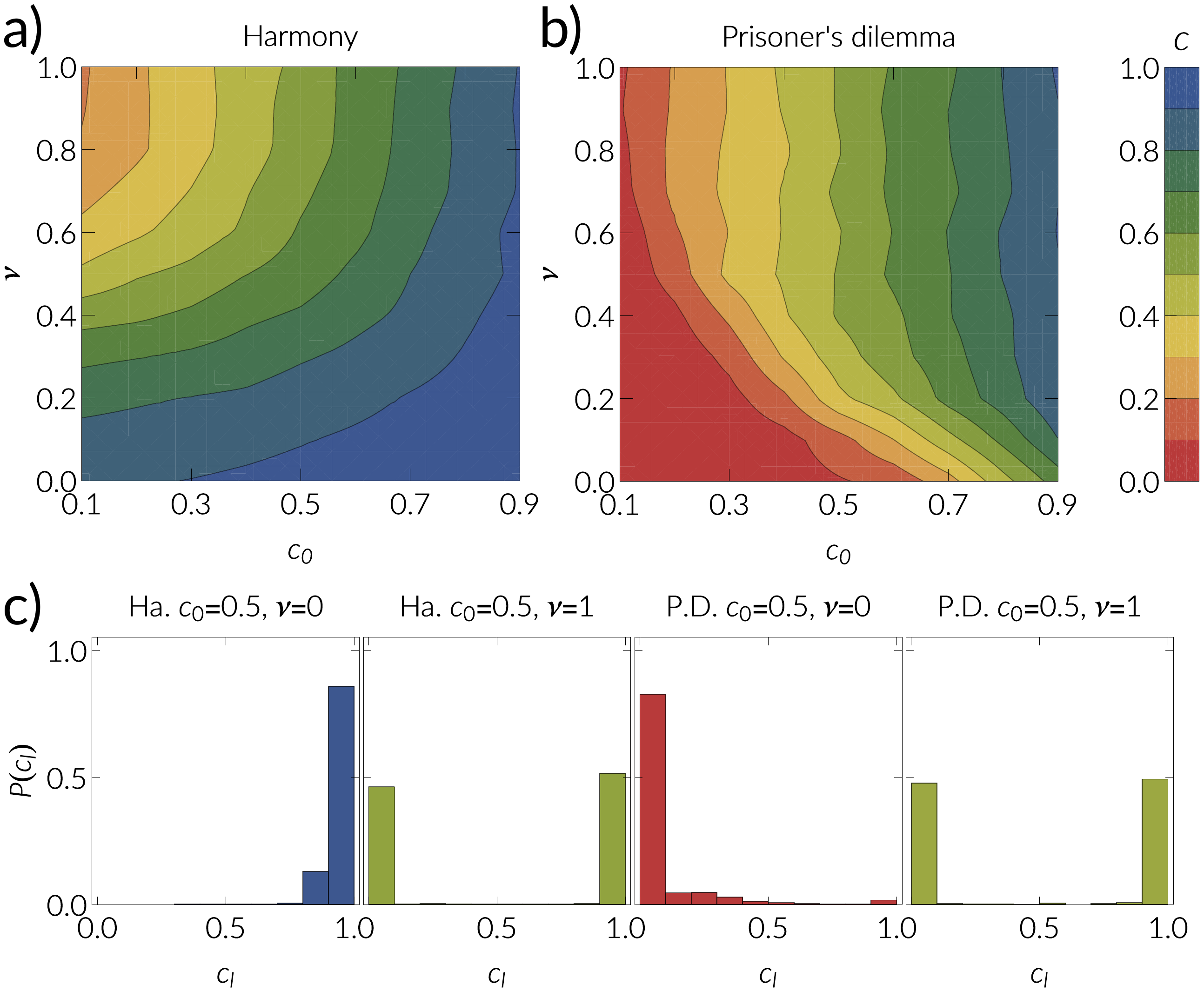}
 \caption{
  Mean final cooperation (color coded) averaged over $400$ realizations for \textbf{(a)} the harmony ($T=0.5$ and $S=0.5$) and \textbf{(b)} the prisoner's dilemma ($T=1.5$ and $S=-0.5$) game as a function of the inital density of cooperators $c_0$ and the strength of radial correlations $\nu$.
   Multiplexes are generated with the geometric multiplex model and have $16$ layers and are generated with the model described in the text. Layers have $N=4000$ nodes, power-law exponents $\gamma=2.6$, and mean local clustering coefficient $0.4$. 
   \textbf{(c)} Histograms of the cooperation density $c_l$ in different layers $l$. We show results for the uncorrelated as well as the correlated case for the harmony game and prisoner's dilemma as indicated in the plot titles.
 \label{fig:InitNu}}
\end{figure}

Moreover, topological enslavement implies that the initial conditions play an important role for the final outcome. If we start for example with $30\%$ cooperators, the hubs will also be initially cooperative with $30\%$ probability and, following the aforementioned mechanism, on average $30\%$ of the layers will become cooperative, whereas the remaining $70\%$ will become defective. 
Indeed, 
we observe that initial conditions determine the outcome if degree correlations are strong. 
We show this in Fig.~\ref{fig:InitNu}, where we vary the initial cooperation $c_0$ from $10\%$ to $90\%$. 
For weak degree correlations, in the harmony game we observe mainly cooperation, and in the prisoner's dilemma mainly defection. 
However, as the strength of degree correlations increase, the final outcome aligns very well with the initial density of cooperators, approximatively independently of the game that is played. 

Finally, the numerical results confirm that topological enslavement separates the layers into two groups with either near full cooperation or defection respectively. This behavior is shown in Fig.~\ref{fig:InitNu}c, where we have set $c_0=0.5$. In the absence of degree correlations ($\nu=0$), in the harmony game we find that cooperation is high in all layers, whereas in the prisoner's dilemma defection prevails for the parameters $T$ and $S$ considered here. 
However, if degree correlations are strong ($\nu=1$), we observe that in both games layers are either highly cooperative or nearly fully defective. The probability of a random chosen layer to be highly cooperative is then approximatively equal to $c_0$, and hence in this case $50\%$. In the Supplementary Materials we show this behavior for different values of $c_0$ and $\nu$.

\section{Discussion}
To conclude, humans constantly interact in different domains where they can adopt different strategies in evolutionary games. 
A player who interacts with an opponent in one domain may not have knowledge of the opponent's behavior in the other domains. 
Therefore, the success of the opponent in terms of her aggregated payoff can be the result of interactions in other domains, where she may have played a different strategy.
Because these interactions are ``hidden'' from the player, she will attribute the success of the opponent to her strategy in the domain where the two interact.
We have shown that the degree of cooperation cannot be modulated by the payoff parameters of the game if the system meets certain topological conditions, which are found in most real multilayer systems. 
This means that payoff-based incentive schemes can become ineffective in environments with multiple domains of interaction.

In particular, the different domains can be represented as a multiplex network. In these type of systems, individuals are simultaneously present in different network layers with different topologies. 
In reality, the topologies of different layers are often related, commonly featuring a large edge overlap, degree correlations, and correlations between the similarity of nodes. 
We have shown that if degree correlations between different heterogeneous layers
are strong enough and individuals engage in many domains, incentives that represent the payoff in strategical games can fail. In this case, the final outcome is strongly determined by the initial conditions. 
This phenomenon, which we call topological enslavement, occurs because nodes that are hubs in different layers can accumulate a high aggregated payoff such that other nodes will tend to imitate their strategies in a specific domain, regardless of whether their payoff was earned in this domain. Hence, if individuals interact in multiple domains in a way that imitates more successful behavior, the payoff of strategical games can become irrelevant for the final outcome.

The resulting situation can be interpreted as an encrusted society, which is insensitive to payoff-based incentives.
Mixing the influence individuals have in different domains destroys the correlations between the layers and hence is an effective measure to avoid this situation, but might not be practical in reality. 
Therefore, it constitutes an important task for future research to investigate whether the transition from such payoff-based incentives, where one incentivizes ``how to act'', to topological incentives, which promote ``with whom to act'', could provide a cure for an encrusted society. 
Finally, it would be interesting to conduct experiments with human subjects to verify our theoretical results.

\begin{acknowledgments}
This work was partially funded by the European Community’s H2020 Program under the funding scheme “FET-PROACT-1-2014: Global Systems Science (GSS)”, grant agreement 641191 “CIMPLEX: Bringing CItizens, Models and Data together in Participatory, Interactive SociaL EXploratories”
and partially by SNF (Grant No. 162776, ``The anatomy of systemic financial risk'') (K.-K.K.). 
D.H. is grateful for support by the ERC Grant “Momentum” (Grant No. 324247).
\end{acknowledgments}

\end{document}